\documentclass{mem}
\usepackage{mybib}\usepackage{txfonts}\usepackage{balance}
\usepackage{graphicx}

\idline{22}{233}

\begin{document}
\title{Pre-MS depletion,  accretion and primordial $^7$Li }

\author{
P. \,Molaro\inst{1} 
\and A. \, Bressan\inst{2}
\and M. \, Barbieri\inst{3}
\and P. \, Marigo\inst{4}
\and S. \, Zaggia\inst{3}
          }

\offprints{P. Molaro}

\institute{
INAF-Osservatorio Astronomico di Trieste, Via G.B. Tiepolo 11, I 34143,  Trieste, Italy
\and
SISSA, Via Bonomea 265, 34136 Trieste, Italy
\and
 INAF-Osservatorio Astronomico di Padova, Vicolo dell'Osservatorio  5, 35122 Padova, Italy
 \and
Dipartimento di Fisica e Astronomia. Universit\'a degli Studi di Padova, Vicolo dell'Osservatorio  5, 35122 Padova, Italy
}

\authorrunning{Molaro et al.}

\titlerunning{Pre-MS Li evolution}

\abstract{
    We reconsider the role of pre-main sequence (pre-MS) Li depletion on the basis of  new  observational and theoretical evidence:   i) new observations of H$\alpha$ emissions in young clusters  show  that mass accretion could be continuing  till  the  first stages of the MS,  ii) theoretical implications from helioseismology 
suggest large overshooting values below the bottom of the convective
envelopes. We argue here that a significant pre-MS $^7$Li destruction, 
caused by efficient overshoot mixing, could be followed by a matter accretion after $^7$Li depletion has ceased
on MS thus restoring Li almost to the pristine value.
   As a test case we show that a  halo dwarf of     0.85  $M_{\odot}$  with an extended overshooting envelope     starting with an initial abundance of  A$(Li) =  2.74$  would burn Li completely, but      an  accretion rate of the type $  1 \times 10^{-8} e^{-t/3\times 10^6}$   $M_{\odot}$  yr$^{-1}$  would restore Li  to end with an A$(Li) =  2.31$.  A self-regulating process is required  to produce similar final values in a range of different stellar masses to explain the PopII Spite plateau.    However, this framework could explain  why  open cluster stars  have  lower Li  abundances   than  the   pre-solar nebula , the absence of Li in the  most metal poor dwarfs    and a number of other features  which lack of a  satisfactory  explanation.
}
 \maketitle{}

\section{Introduction}

\begin{figure}[t!]
\resizebox{\hsize}{!}{\includegraphics[clip=true,angle=-90]{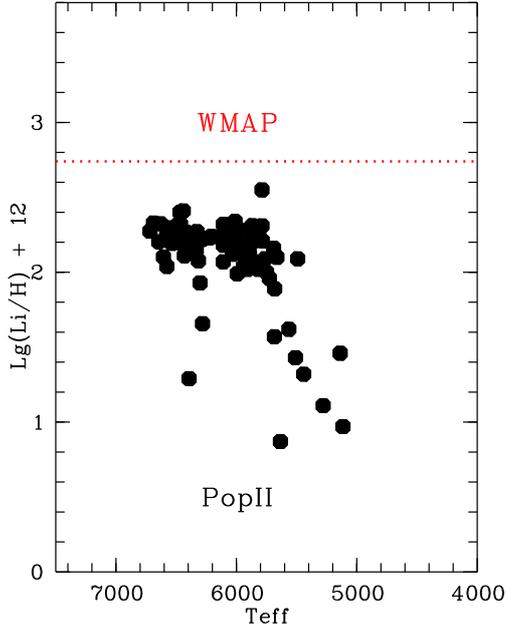}}
\caption{\footnotesize
 Li observations in PopII stars. The red line marks the  WMAP Li prediction}
\label{label:fig_pop2}
\end{figure}
\begin{figure}[t!]
\resizebox{\hsize}{!}{\includegraphics[clip=true,angle=-90]{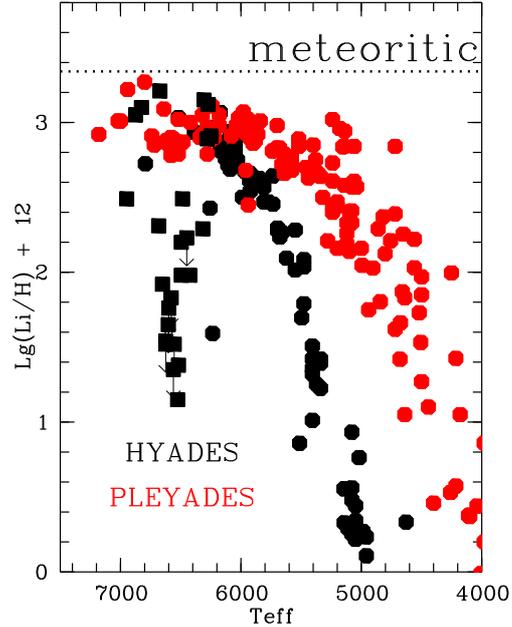}}
\caption{\footnotesize
Li observations in Hyades, black dots and squares, and Pleiades, red dots. The dotted line shows the level of meteoritic $^7$Li abundance
}
\label{label:fig_clusters}
\end{figure}

\subsection{The PopII-WMAP puzzle}
  
 The WMAP  value for $\Omega_b h^2$  yields 
    $\eta_{10} = 10^{10} (n_{b}/n_{\gamma})_0$ =   $6.16 $ (Komatsu et al 2011),  light elements in the framework of SBBN.   
  The predicted primordial $^7$Li abundance is   A(Li)=2.72 , in the scale where A(Li)=$\log[N(Li)/N(H)]+12$,  (Coc et al 2012).
 
On the other hand, almost 3 decades of  Li  observations in   PopII stars provided  an abundance of   $A$(Li)$\approx$ 2.26 (cfr Molaro 2008),   missing the WMAP prediction by about a factor 3 as emphasized in Fig \ref{label:fig_pop2}. The data in the figure are drawn from  the 1D non-LTE abundances with IRFM temperatures from Bonifacio and Molaro (1997) and Sbordone  et al (2010).

The solution of the Li problem was widely debated at this conference. Very different approaches have been  considered  from the nuclear reaction rates  in the SBBN (Coc et al 2012) to   new physics beyond the standard model (Olive 2012, Kajino 2012) or  depletion in stars by diffusion (Richard 2012, Korn 2012).  
\subsection{The PopI-meteoritic puzzle}
The PopII-WMAP  problem is mirrored at high metallicity by the difference between the meteoritic $^7$Li abundance and that  observed in the open clusters.
The present $^7$Li abundance at the time of the formation of  the solar system as    obtained from meteorites is $A$(Li)$=3.34 \pm0.02$ (Anders and Grevesse 1989). For recently born stars, the initial $^7$Li abundance has been  estimated in young T-Tauri stars where  no Li
depletion could  yet have taken place A$(Li)\, =  3.2 \pm 0.1$  (Cunha et al 1995). The hotter F stars of
slightly older clusters such as  Pleiades  ( $\approx$ 100  Myrs ) and Hyades (670 Myrs)  show a  top value of  
  $A$(Li)$\approx 3.0$,  Thorburn 1994, Soderblom 1993)  and never reach the  meteoritic value as shown in Fig \ref{label:fig_clusters}.  
 The behavior of several other young clusters is quite similar to the Pleyades  but the data are omitted in the figure for clarity (Sestito and Randich 2005).  Considering that the present Li abundance  could be higher than the meteoritic value  due to Li Galactic  enrichment,    the disagreement  between PopII and WMAP and cluster plateau's and meteoritic values could be considered  of  comparable scale. We  examine here the possibility that  they may share  a common origin dealing with  the pre-MS $^7$Li evolution.

\section{Pre-MS Li evolution}

 Pre-MS stars with masses $M<0.5\,M_{\odot}$   are almost 
fully convective  all the way to the ZAMS and  as the star
contracts along its Hayashi track, its core heats up.
 When the core
temperature  $T_{c}$ reaches $\simeq 3\times10^{6}$\,K,   Li begins to burn in $p,\alpha$ reactions. These reactions
are  very  temperature-sensitive ($\propto T_{c}^{16-19}$)   and convective mixing so fast  that  all   $^7$Li is burned in
a few Myr. 

In 
1\,$M_{\odot}$ stars photospheric Li depletion begins at about 2\,Myr and  terminate at about 15\,Myr. This range
shifts towards older ages in lower mass stars and Li-burning temperatures are never reached for
$M<0.06\,M_{\odot}$. 

\begin{figure*}
\resizebox{\hsize}{!}{\includegraphics[clip=true]{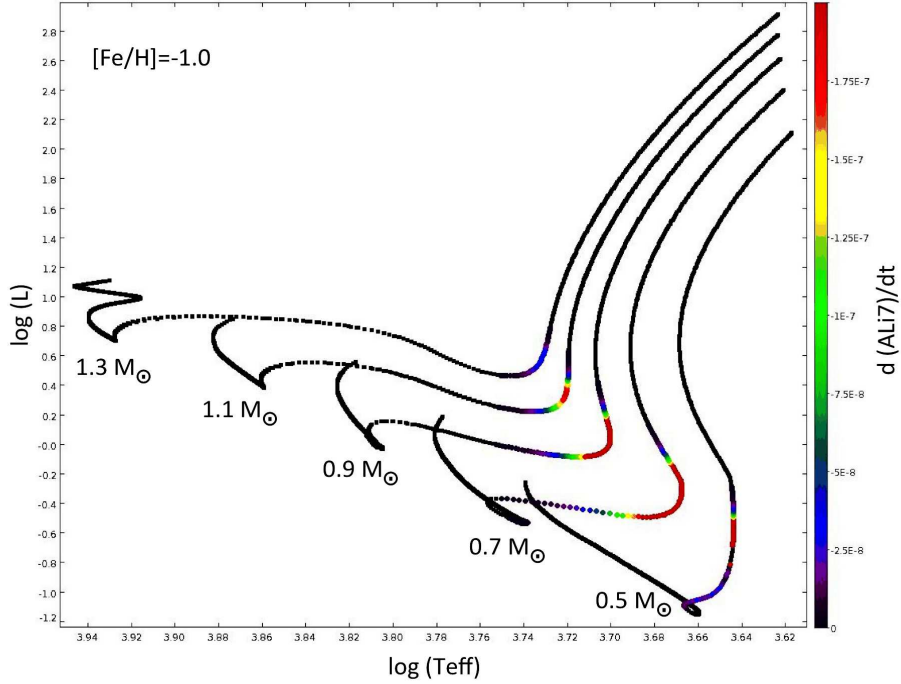}}
\caption{\footnotesize
Locus in the HR diagram corresponding to the Li burning. The red segment shows the maximum burning.
}
\label{label:fig_hr}
\end{figure*}

  Li depletion is therefore
a sensitive function of age between about 10 and 200\,Myr. In PMS stars luminosity, $L \propto M^{2}$  and in Fig \ref{label:fig_hr} the locus  of the HR diagram corresponding to the Li burning is shown. The red segments show the luminosities for the maximum burning of different masses.

The overall amount of Li-depletion is also extremely sensitive to mass and there should be relatively little depletion
in solar mass stars if compared with lower-mass stars. For
$M<0.6\,M_{\odot}$ all the Li is burned. 
For higher mass stars the radiative core
develops before Li burning is complete and the temperature at the base
of the convective envelope, $T_{bcz}$, decreases. 
In the absence of
convective mixing, Li-depleted material cannot get to the photosphere,
so once $T_{bcz}$ drops below the Li-burning threshold, photospheric
Li-depletion ceases.

Photospheric
Li-depletion arises from rapid Li burning in a very thin region above
the convection zone base, and current models  are  too coarse to predict Li
depletion accurately. The exact amount of expected Li depletion  is  dependent on a
number of model details such as  the time at which the radiative core develops, the position of
the convection zone base and hence $T_{bcz}$. As a result,  large changes
in Li depletion predictions can result from relatively minor
perturbations in model parameters. As discussed in Ventura et al (1998) and also Tognelli et al at this conference,  there is a  unsatisfactory agreement between models and the observational data. However, the models are generally made to fit the Li abundances of the hotter stars in the open clusters.
Convective efficiency and overshooting are crucial model parameters which determine the extension of the convection zone. If overshooting is
present,   then the temperature at the base of the convective zone is higher  resulting in much more
photospheric $^7$Li depletion.

A recent analysis of 
helioseismology data
(Christensen-Dalsgaard et al 2011)
indicates that external convection in our Sun
could penetrate into the underlying stable regions
by a significant amount, corresponding to
an overshoot scale $\Lambda_{\rm OV}\sim 0.4 \sim 0.6 H_P$.
These values indicate that overshoot mixing 
could be much more efficient than previously 
believed.

In the following we will assume that the efficiency of the
overshoot process during the PMS phase is  large enough
to allow an almost complete depletion of $^7$Li.
For the  test case presented here (Fig 4) we consider
a star with 0.85~M$_\odot$,
[Fe/H]=-2.2, initial A(Li)=2.74
and an overshooting parameter $\Lambda_{\rm OV}=1.5H_P$.
The latter is not unreasonable
since, during the PMS phase, a 
significant fraction of the star is convective.

\balance
\section{Pre-MS LI evolution with mass accretion}

Recent observations  of  H$\alpha$ emissions in several fields  suggest that accretion could last longer than formerly believed  (De Marchi et al 2010, 2011, Spezzi et al 2011). It is noteworthy that the majority of PMS objects showing evidence of  matter accretion 
are very  close to the MS. The median mass accretion rate they derive on the MS is of 
 $\dot M_{\rm acc}$ = $2.6
\times 10^{-8}  M_{\odot}$ yr$^{-1}$. 

We note that this is  gas  and not  planetesimal  accretion. The latter   is  H/He depleted and in order  to explain the Li abundance scatter this would also
lead to  Fe abundance anomalies of the order of 0.2-0.3\,dex --
much higher than allowed by current observational constraints.

 If accretion occurs only during the initial phases of pre-MS evolution,  the new Li
brought in by accretion is depleted as the initial one. 


However, the Li evolution would be completely different if the accretion takes place  also  {\em after}  the Li depletion phase. In fact, 
accretion occurring  after Li-burning has ceased would enrich  again the
convective zone  with Li.  

The convective zone masses  for these stars are  
  few  percent of their mass. They  are    $M_{CZ}= 0.01- 0.06\,M_{\odot}$, for 1 to 0.7 $M_{\odot}$,  respectively. The  accreted material with pristine$^7$Li is rapidly diluted in the convection zone (CZ) where $^7$Li has been previously burnt and  even  a relatively small mass accretion has  a non-negligible impact on the resulting Li abundance.

    We  modified our stellar evolution code
(Bressan et al 2012) in order to account for
low accretion rates.
In our test case we consider that the accretion rates
declines as    $\dot M_{\rm acc}$ =      $  1 \times 10^{-8} e^{-t/t_0 }$   $M_{\odot}$  yr$^{-1}$  where t is in yr and t$_0$ = 3 $\times 10^6 $yr.

The accreted material is mixed through the convective zone
and eventually burned by nuclear reactions in the star center 
or at the bottom of the convective zone if the temperature is high 
enough.

Our results do not depend on the previous  evolution
along the stellar birth-line (Stahler, 1983) because,
when  large accretion ends,
deuterium burning cannot more be sustained and  the star
exits from the stellar birth line 
(blue dashed line in the HR diagram of Fig. 4) 
and evolves along the almost constant mass path (red line).
Soon after, deuterium is almost completely exhausted. This phase is marked with a triangle in the same figure.

The time behavior of surface element abundances, central temperature,
temperature at the bottom of the convective envelope and
mass accretion rate are all shown in the right panel of Fig. 4.
Deuterium is almost immediately destroyed and at the base of 
the Hayashi track also $^7$Li starts to be  depleted.
Due to the large overshoot it is almost completely
depleted after about 4 Myr since Deuterium depletion.
At this stage the central region becomes  radiative and 
convection starts receding towards the external layers. 
The temperature at the base of the convective envelope
falls below the $^7$Li burning limit and eventually also 
below the deuterium burning limit.
Here accretion is $\sim$10$^{-9}$ $M_{\odot}$ yr$^{-1}$
but, since the convective region gets smaller and smaller, this
rate is high enough to restore high values of the
surface $^7$Li abundance. Incidentally surface Deuterium rises to 
almost the pristine value but then it is burned again
and destroyed. 
With the parameters adopted in this test
case the surface $^7$Li stabilizes at A(Li)=2.31, i.e. close to
what observed in the PopII stars.


 \begin{figure*}
\resizebox{\hsize}{!}{\includegraphics[clip=true,angle=90]{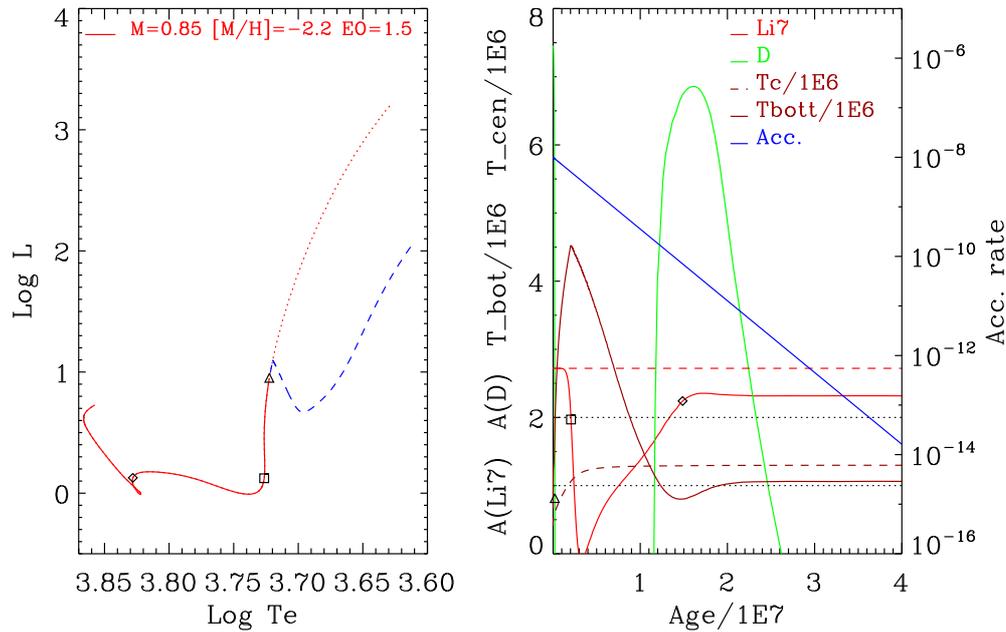}}
\caption{\footnotesize
Left panel: pre MS evolution of a star with 0.85  $M_{\odot}$, with [Fe/H]=-2.2 and an overshooting envelope of 1.5. The dotted blu line is the evolution of a star with initial mass
 of 0.2  $M_{\odot}$ accreting mass at a rate of  $\sim$10$^{-5}$ $M_{\odot}$ yr$^{-1}$ till it becomes a star of 0.85 $M_{\odot}$. Since then the accretion is reduced to $\sim$10$^{-8}$ $M_{\odot}$ yr$^{-1}$. The empty square marks the onset of Li burning and the diamond  the end of the Li accretion.  The triangle marks the end of the D burning and of the birthline track.  Right panel: Li evolution from an initial A(Li)=2.7 (red line) , an accretion rate with the law given in the text  (blue line). 
}
\label{label:label}
\end{figure*}


\bigskip

\section{Conclusions}

Pre-MS Li depletion is presently  considered to explain the scatter and the depletion of the low temperature side of  Li diagrams in young clusters but thought to be minimal for the hotter members of  cluster members (Tognelli et al 2012).  It is generally considered absent in  PopII dwarfs.  D'Antona \& Mazzitelli (1984)  allow a rather small amount of Li depletion in pre-MS  of the order of 0.15 dex. Instead,  we argue for an effective Li  burning along the pre-MS phase. The  apparent contrast with the observations  which do not show   clear evidence for pre-MS  depletion could be removed if the destruction is substantially balanced by  Li accretion occurring in the late stages of the pre-MS evolution after  Li has been burned in the stellar convective zones. We have shown  a test case of  a 0.85  $M_{\odot}$  PopII dwarf with [Fe/H]=-2-2 with  an original Li at the WMAP value. A   self-regulating process, not yet identified, should be at work in order to  produce  almost the  same dilution factor  for the variety of stellar masses  which are on  the Spite plateau or in the hotter stars of the open clusters. The onset of the  stellar wind could counteract the lithium restoration both by  removing  the outer layers and also by dissipating  the accreting disk.  This is  a working hypothesis but  we note that  it has several  specific assets, namely:

\begin{itemize}

\item{Providing a possible explanation of the disagreement  between PopII Li abundances and WMAP predictions.}

\item{Providing a possible explanation of why models with no pre MS Li depletion fail in explaining open clusters Li diagrams and in particular of why  F stars Li abundances are always below the  solar-meteoritic  value.} 
\item{Failure  or increase of the accretion process could provide an explanation for the few Pop II  $^7$Li depleted stars ( $\approx$ 3\%)   or of the   few  Pop II star showing a   $^7$Li abundances at the WMAP level, respectively.  Among the latter there is  BD +23 3912  with  $A(Li)$=2.60 which  stands out the others in  Fig \ref{label:fig_pop2}(Bonifacio and Molaro 1997).}
\item{ The extreme metal poor dwarfs  of Caffau et al (2011) and Frebel et al   (2005) without detectable Li 
could have been formed in a fragmentation process  and never  accreted material after a complete $^7$Li depletion in the pre-MS phase. Caffau et al (2012) by using a set of different  isochrones to match the observed colors estimate  a mass  in the range   0.62-0.70  $M_{\odot}$. Since for
$M<0.6\,M_{\odot}$ all the Li is burned in pre-MS, the absence of Li could be explained by an extended pre-MS depletion in a star with an initial mass below this threshold. In fact the absence of Li in these stars can be considered as evidence  of pre-MS Li depletion in  extreme metal poor stars. The  melting  of Li abundances   for  [Fe/H] $\le$ -3.0 observed by Sbordone et al (2010) could be another feature of the mass decrease at lower metallicity for a given effective temperature.  Towards the low metallicity tail  stars have progressively smaller masses and the  mass accretion could be not enough to restore Li at the level of the more massive stars. }

\item{It could provide an explanation of the Li paradox in the metal poor spectroscopic binaries.  The    PopII spectroscopic binaries  CS 22876-032 (Gonzales Hernandez et al 2008) and  G 166-45  (Aoki et al 2012) are composed by dwarfs with a temperature characteristic of the plateau  but  show slightly different  Li  abundances.  The hotter  is on the Spite Plateau but  the cooler show a lower  Li abundance  of  $\approx$ 0.4 dex and $\approx$ 0.2 dex respectively. The  masses of the CS 22876-032 system are $m_1 =0.70 \pm 0.02$ and $m_2 = 0.64 \pm 0.02$  and those of the G166-45 system are  $m_1 =0.76 \pm 0.02$ and $m_2 = 0.67 \pm 0.02$. The cooler has also a smaller mass and very close to the full convective limit.  Thus it is quite possible that  with the same accreted mass the restoring of  Li has  been lower in the smaller mass star due to the relatively larger convection zone.}

\end{itemize}

\begin{acknowledgements}
It is a pleasure to thank Giacomo Beccari, Lorenzo Monaco, Sofia Randich for useful discussions and  Gabriella Schiulaz for checking the  English text.
\end{acknowledgements}

\bibliographystyle{aa}

\end{document}